\documentclass[reprint,twocolumn,superscriptaddress,notitlepage,nofootinbib]{revtex4-2}
\usepackage[T1]{fontenc}
\usepackage[utf8]{inputenc}
\usepackage[english]{babel}
\usepackage[hidelinks]{hyperref}
\usepackage{microtype,amsfonts,amssymb,amstext,amsmath,mathtools,physics}

\begin{document}
\title{Uncertainty Principles and Non-local Black Holes}

\author{Salvatore Capozziello}
    \email{capozziello@na.infn.it}
    \affiliation{Dipartimento di Fisica ``E.\ Pancini'', Università di Napoli ``Federico II'', Via Cintxhia 9, 80126 Napoli, Italy}
    \affiliation{Scuola Superiore Meridionale, Largo San Marcellino 10, 80138, Napoli, Italy}
    \affiliation{INFN Sezione di Napoli, Complesso Universitario di Monte Sant'Angelo, Edificio 6, Via Cinthia, 80126, Napoli, Italy}

\author{Giuseppe Meluccio}
    \email{giuseppe.meluccio-ssm@unina.it}
    \affiliation{Scuola Superiore Meridionale, Largo San Marcellino 10, 80138, Napoli, Italy}
    \affiliation{INFN Sezione di Napoli, Complesso Universitario di Monte Sant'Angelo, Edificio 6, Via Cinthia, 80126, Napoli, Italy}

\author{Jonas Mureika}
    \email{jmureika@lmu.edu}
    \affiliation{Department of Physics, Loyola Marymount University, Los Angeles, California, USA}

\date{\today}

\begin{abstract}
    We discuss the Generalized Uncertainty Principle and the Extended Uncertainty Principle in the context of black hole solutions coming from non-local theories of gravity, focusing, specifically,  on Infinite Derivative Gravity. We argue that these modifications of the Heisenberg Uncertainty Principle are effective descriptions arising from the non-local features of  gravitational interaction. By comparing the predictions of both the modified uncertainty principles and non-local gravity, we find theoretical constraints on otherwise free parameters as well as universal laws for black hole physics beyond General Relativity.
\end{abstract}

\maketitle

\section{Introduction}
Black holes represent the main theoretical arena used to investigate deviations from both the established theory of matter, that is Quantum Mechanics (QM), and that of gravity, that is General Relativity (GR). As such, black holes are of paramount interest for any Quantum Gravity (QG) research program. A promising avenue for solving some of GR's shortcomings is related to non-locality, which can be introduced into a theory of gravity either as an effective description or as prompted by first principles. Infinite Derivative Gravity (IDG) is a broad set of non-local theories of gravity developed on the basis of strong theoretical motivations, which result in the formulation of singularity-free, unitary and renormalizable theories of QG (see e.g.\ the Refs.\ \cite{biswas:towards,biswas:bouncing,modesto:super}). IDG theories take into account analytic functions of the generally covariant d'Alembert operator or its inverse, leading to ultraviolet (UV) or infrared (IR) corrections to GR respectively. 

Assuming terms involving the Ricci scalar only, an effective IR non-local gravitational action for IDG can be written as (see Ref.\ \cite{capozziello:really} and references therein)
\begin{equation}
    S=\int d^{4}x\,\frac{\sqrt{-g}}{2}[M_\textup{P}^2R+R\bar{\mathcal{F}}_1(\Box)R],
\end{equation}
where
\begin{equation}
    \bar{\mathcal{F}}_1(\Box)=\sum_{n=1}^\infty f_{1_{-n}}\Box^{-n},\qquad f_{1_{-n}}\equiv\tilde{f}_{1_{-n}}M_\textup{IR}^{2n}.
\end{equation}
Here $\tilde{f}_{1_{-n}}$ are dimensionless constants; $M_\textup{P}=2.44\cdot10^{18}$\,GeV is the reduced Planck mass and $M_\textup{IR}$ is an IR mass scale. In the UV limit, $M_\textup{IR}$ can be neglected with respect to $M_\textup{P}$ and the action reduces to the Einstein--Hilbert action of GR.
The inverse d'Alembert operator $\Box^{-1}$ can be expressed in terms of its Green function and then solved. On the other hand, the  action
\begin{equation}
    S=\int d^{4}x\,\sqrt{-g}F(R,\Box^{-1}R),
\end{equation}
where $F$ is a generic analytic function, can be considered to  take into account both UV and IR corrections \cite{bajardi:non-local}. UV non-local modifications seek to resolve singularity problems like that of black holes or the Big Bang \cite{biswas:bouncing,modesto:super,briscese:inflation,biswas:stable,modesto:complete}, while IR non-local modifications aim at explaining mainly dark matter or dark energy \cite{deser:nonlocal1,deser:nonlocal2,arkani-hamed:non-local,barvinsky:nonlocal,nojiri:screening,nojiri:ghost-free,elizalde:desitter,bamba:screening,lev:solving,capozziello:reconstructing,maggiore:phantom,foffa:cosmological,maggiore:nonlocal,capozziello:really,capozziello:weinberg,bouche:testing,bouche:addressing,zhang:screening,bombacigno:inflation}.

Effects of  non-locality in  spacetime can manifest in various ways, but, in general, they introduce a fundamental `fuzziness' to the gravitational field related to  GR and a characteristic length, either at very small or very large scales. This concept is similar to that implied by modifications of the Heisenberg Uncertainty Principle, namely the Generalized Uncertainty Principle (GUP) and the Extended Uncertainty Principle (EUP). The former can be derived from a variety of QG theories and affects black hole physics at the Planck scale \cite{kempf:hilbert,adler:gravity,veneziano:stringy,witten:reflections,scardigli:gedanken,gross:string,amati:spacetime,yoneya:interpretation,ashtekar:quantum,hossain:background,isi:self-completeness,majid:scaling,maggiore:generalized,maggiore:algebraic,maggiore:quantum,adler:generalized,chen:remnants,tawfik:generalized,carr:generalized,carr:correspondence,carr:black,koppel:generalized,carr:sub-planckian,amelino-camelia:thermodynamics}, while the latter plays a role at astrophysical or cosmological scales \cite{kempf:hilbert,bambi:natural,mignemi:extended,zhu:influence,mureika:extended,pantig:extended}. In both cases, GUP and EUP provide corrections to the uncertainty which is intrinsic to any measurements made on matter fields. This additional `fuzziness' of matter manifests itself at  characteristic length scales, either in the UV or IR regimes. This extra uncertainty in either the position or momentum space directly translates into an additional non-local behaviour of particles, whose intrinsic delocalisation in spacetime produces observable gravitational effects beyond the classical picture of Einstein gravity. Therefore, QG effects demand a reconsideration of gravitational physics especially in regimes where GR exhibits theoretical deficiencies, like the UV limit of black hole or wormhole solutions and the IR dynamics of dark matter. For example, the GUP occurs in many interesting astrophysical systems which have been extensively studied in the context of modified theories of gravity \cite{mondal:entropy,battista:generalized,rani:casimir,joshi:hamiltonian,malekolkalami:null,ali:entropy,christiansen:new,capozziello:null,capozziello:avoiding}.

In this work we argue that, in the context of black hole physics, GUP and EUP are effective descriptions stemming from  non-locality effects in gravity. This insight can be understood as follows. To begin with, both the paradigm for non-locality and that for  modified uncertainty principles propose the existence of an inherent `fuzziness' or `smearing' in the way matter and spacetime interact, which goes beyond the fundamentals of both QM and GR. Moreover, in both cases, these novel non-local effects are expected to be relevant only at some characteristic length scales; therefore, one can naturally be led to identify the characteristic UV or IR length scales of non-local  gravity with that of GUP or EUP respectively. The underlying reason for the alleged connection between non-locality and modified uncertainty principles ultimately comes from dynamics. On one hand, GUP and EUP assume that gravity is described by the laws of GR and so introduce modifications on the behaviour of matter only; on the other hand, non-local gravity deals with the ordinary matter that obeys the rules of QM, while altering the local nature of GR's spacetime curvature. But modifications of the matter sector must reflect in modifications of the gravitational sector and vice versa, hence why GUP and EUP can be seen as effective descriptions in which only the theory of matter is changed while maintaining the tenet of locality in the description of spacetime. Instead, non-local gravity suggests a radical reformulation of the gravitational interaction which, in some appropriate UV or IR regimes, could be well approximated by the very different theoretical framework of GUP or EUP respectively. The idea of linking non-local black holes and GUP is not new. In Ref.\ \cite{nicolini:nonlocal}, a GUP-inspired model of black holes was found to be equivalent to the one derived from a specific non-local theory of gravity, thus extending previous findings about black hole spacetimes inspired by noncommutative geometry. This result is due exactly to the fact that the attribution of the non-local features can freely be shifted from the matter source to spacetime or vice versa by making use of the gravitational field equations. In the case of some specific models, that work also highlighted some intriguing physical possibilities, such as a zero-temperature remnant formation at the end of the evaporation process and a regular de Sitter core accounting for the energy density of virtual gravitons in place of the curvature singularity. Here we look for more general relations between non-locality and modified uncertainty principles for black holes, thus we consider IDG theories as the starting point of our investigation. Natural units are adopted in the paper.

\section{UV non-local gravity and GUP}
Based on various arguments coming from QG, the GUP hypothesises a momentum-dependent correction to the Heisenberg Uncertainty Principle:
\begin{equation}
    \Delta x\Delta p\geq1+\beta L_\textup{P}^2\Delta p^2,
\end{equation}
where the (reduced) Planck length is defined as $L_\textup{P}\equiv\sqrt{8\pi G}$ and the parameter $\beta$ is dimensionless (in natural units). For a very preliminary study in this direction, see Ref.\ \cite{capozziello:generalized}. A GUP-inspired theory of gravity was proposed in \cite{carr:sub-planckian}, including the prediction of sub-Planckian remnants as dark matter candidates, and further explored in \cite{carr:rnkerr,carr:reconciling}, in which a GUP-modified Schwarzschild metric with the following form was introduced:
\begin{eqnarray}\label{eq:GUPmetric}
	{ds}^2&=&f_\textup{GUP}(r){dt}^2-\frac{{dr}^2}{f_\textup{GUP}(r)}-r^2{d\Omega}^2,\nonumber\\
	f_\textup{GUP}(r)&=&1-\frac{2Gm}{r}\left(1+\frac{\beta}{2}\frac{M_\textup{P}^2}{m^2}\right).
\end{eqnarray}
In principle, the proposed modification could apparently allow for the existence of sub-Planckian black holes with masses $m\ll M_\textup{P}$. However, in the remainder of this section we will argue that non-local gravity can fix the form of the mass-dependent parameter $\beta$ in such a way as to exactly cancel the dependence on mass for the GUP-inspired correction. This leads to the prediction of some universal deviations from GR for astrophysical black holes, while leaving effectively unaltered the conclusion that particle-sized masses cannot be black holes. For some astrophysical investigations of the GUP along these lines, see for example the Refs.\ \cite{feng:constraining,feng:quantum,feng:jeans}.

In the weak field limit, the following Newtonian potential is recovered:
\begin{equation}\label{eq:Phi_GUP}
    \Phi_\textup{GUP}(r)=-\frac{Gm}{r}\biggl(1+\frac{\beta}{16\pi Gm^2}\biggr).
\end{equation}
The corresponding event horizon is thus shifted with respect to that of GR, as
\begin{equation}\label{eq:R_EH1}
    R_\textup{EH}=\biggl(1+\frac{\beta}{16\pi Gm^2}\biggr)R_\textup{S},
\end{equation}
where $R_\textup{S}\equiv2Gm$ is the Schwarzschild radius. Note that the metric \eqref{eq:GUPmetric} exhibits a feature of dimensional reduction in the sub-Planckian regime, yielding black holes with (1+1)-D horizon and thermodynamic characteristics. That is, for $m\ll M_\textup{P}$ the horizon scales as $R_\textup{EH}\sim m^{-1}$ and $T\sim m$, which cures the divergent behaviour of the standard Hawking temperature profile in the limit $m\rightarrow0$. Although in other frameworks black holes of this type could be subsequently large for particle-sized masses, this does not imply that particles themselves are black holes in this case. Despite the functionally-similar form of the (lower-dimensional) horizon radius and the Compton wavelength, in fact, there is no reason to suggest that the large-mass Schwarzschild behaviour flows smoothly into the Compton relation, due to the different scaling nature of each. The interested reader is referred to Ref.\ \cite{carr:sub-planckian} for further reading on this topic, as well as to Ref.\ \cite{carr:rnkerr} for a discussion on the alleged self-completeness of the gravitational interaction. Similarly, the nature of the potential \eqref{eq:Phi_GUP} for particle-sized masses was extensively explored in Ref.\ \cite{carr:reconciling}, {including experimental constraints.

A common feature of UV non-local theories of gravity is the effective `smearing' of point-like sources over a region whose size is defined by the characteristic length scale of the theory. Within this region, the effects of non-locality become significant, resulting in a regular rather than singular gravitational field. To see this, one can consider a sub-class of UV non-local theories with infinitely many d'Alembert operators \cite{biswas:quadratic} in the weak-field limit. For a static point-like source of mass $m$ and energy density $\rho$, the non-local field equations yield the following modification of Poisson's equation \cite{buoninfante:classical} with respect to GR:
\begin{equation}
    e^{-L_\textup{UV}^2\Box}\nabla^2\Phi_\textup{NL}(r)=4\pi G\rho(r),
\end{equation}
where $\Phi_\textup{NL}$ is the spherically symmetric Newtonian potential of the theory, $L_\textup{UV}$ is the characteristic length scale of non-locality and $\Box=\nabla^2$ under the assumption of the matter source being static. By shifting the non-local exponential factor to the r.h.s.\ and employing the Fourier transform representation, this equation becomes equivalent to Poisson's equation of GR for a source `smeared' over a region of finite size $\sim L_\textup{UV}$:
\begin{equation}
    \nabla^2\Phi_\textup{NL}(r)=4\pi Gme^{L_\textup{UV}^2\Box}\delta^3(\vec{r})=\frac{Gm}{2\sqrt{\pi}L_\textup{UV}^3}\exp(-\frac{\vec{r}^2}{4L_\textup{UV}^2}).
\end{equation}
Solving for $\Phi_\textup{NL}$, one finds that
\begin{equation}\label{eq:Phi_UV}
    \Phi_\textup{NL}(r)=-\frac{Gm}{r}\erf\biggl(\frac{r}{2L_\textup{UV}}\biggr).
\end{equation}
This Newtonian potential is asymptotically flat and approximates GR's Newtonian potential for $r\gg L_\textup{UV}$, while, for $r\rightarrow0$, it converges to the finite value $-\frac{Gm}{\sqrt{\pi}L_\textup{UV}}$, thus making the theory asymptotically safe.

The potentials \eqref{eq:Phi_GUP} and \eqref{eq:Phi_UV} are very different, with $\Phi_\textup{NL}$ being structurally different from the Newtonian potential of GR and providing different physical predictions for $r\lesssim L_\textup{UV}$. Still, the potential $\Phi_\textup{GUP}$ can be expected to be an effective description of $\Phi_\textup{NL}$ at some adequate length scale (though not necessarily at all scales). We argue that this length scale is the one at which non-local gravitational effects emerge, i.e.\ $L_\textup{UV}$. To understand this claim and test its predictions, one can then study the equation
\begin{equation}
    \Phi_\textup{GUP}(L_\textup{UV})=\Phi_\textup{NL}(L_\textup{UV}).
\end{equation}
Solving for $\beta$ yields
\begin{equation}\label{eq:beta}
    \beta=-16\pi Gm^2[1-\erf(1/2)]\approx-24.1\,Gm^2\approx-\frac{m^2}{M_\textup{P}^2}.
\end{equation}
We observe two important features of this result: $\beta$ is predicted to be negative and to depend quadratically on the ratio $m/M_\textup{P}$, where $M_\textup{P}=1/L_\textup{P}$ is the (reduced) Planck mass. Substituting this expression for $\beta$ into that for the event horizon \eqref{eq:R_EH1} then yields
\begin{equation}
    R_\textup{EH}=2\erf(1/2)Gm\approx0.5\,R_\textup{S}.
\end{equation}
A further consequence is a modified prediction for the evaporation temperature of the black hole:
\begin{equation}
    T_e\approx2\,T_\textup{H},
\end{equation}
where $T_\textup{H}\equiv1/8\pi Gm$ is the Hawking temperature for GR.

\section{IR non-local gravity and EUP}
Taking inspiration from the discussion of the previous section, we can now try to reproduce the analysis for the IR regime.

The EUP introduces a position-dependent modification of the Heisenberg Uncertainty Principle:
\begin{equation}
    \Delta x\Delta p\geq1+\alpha\frac{\Delta x^2}{L_*^2},
\end{equation}
where $\alpha$ is another dimensionless parameter, while $L_*$ is a large length scale, possibly the Hubble length. Following the formalism for the GUP-corrected metric outlined in \cite{carr:sub-planckian}, a corresponding EUP-inspired theory of gravity was proposed in \cite{mureika:extended}. In this case, the modified metric function is
\begin{equation}\label{eq:EUPmetric}
	f_\textup{EUP}(r)=1-\frac{2Gm}{r}\left(1+\frac{4\alpha G^2m^2}{L_*^2}\right).
\end{equation}
This theory modifies GR for masses whose Schwarzschild radius exceeds the EUP scale, $R_\textup{S}>L_*$, such that a black hole's horizon radius scales as $R_\textup{EH}\sim m^3$. It thus follows that supermassive black holes are natural test objects for the corresponding deviations from GR. A variety of observational consequences are considered in \cite{mureika:extended}, including modifications to photon and matter orbits. Subsequent studies have analysed the EUP gravity's influence on Mercury's perihelion shift, the Shapiro time delay and the precession of S2 star orbits \cite{Okcu:2022sio}, strong lensing effects for Sgr A* and M87 \cite{luxie}, and an EUP emergent Hubble tension \cite{Nozari:2024wir}. Furthermore, the EUP length scale $L_*$ may set the value of the cosmological constant \cite{Pantig:2024asu}.

In this case, the Newtonian potential for a static black hole of mass $m$, in the weak-field limit, is
\begin{equation}\label{eq:Phi_EUP}
    \Phi_\textup{EUP}(r)=-\frac{Gm}{r}\biggl(1+\frac{4\alpha G^2m^2}{L_*^2}\biggr).
\end{equation}
The corresponding event horizon, again, is shifted with respect to that of GR:
\begin{equation}\label{eq:R_EH2}
    R_\textup{EH}=\biggl(1+\frac{4\alpha G^2m^2}{L_*^2}\biggr)R_\textup{S}.
\end{equation}
Note that the potential \eqref{eq:Phi_EUP} is greater than the standard Newtonian case for a mass distribution whose Schwarzschild radius is $R_\textup{S}>L_*$. Here, however, the mass does not need to be a black hole and the Schwarzschild radius merely acts as an internal parameter to the theory. In the case of galaxies, the transition to the EUP potential may coincide with the flattening of the rotation curve \cite{mureika:extended}.

As for non-local gravity, one can consider a sub-class of IR non-local theories with infinitely many inverse d'Alembert operators \cite{conroy:infrared} with a characteristic length scale $L_\textup{IR}$. The spherically symmetric Newtonian potential of the theory for a static black hole of mass $m$ in the weak field limit is
\begin{multline}\label{eq:Phi_IR}
    \Phi_\textup{NL}(r)=-\frac{Gm}{r}\biggl[\frac{1}{2}{}_0F_2\biggl(;\frac{1}{2},1;\frac{r^2}{4L_\textup{IR}^2}\biggr)\\
    -\frac{r}{\sqrt{\pi}L_\textup{IR}}{}_0F_2\biggl(;\frac{3}{2},\frac{3}{2};\frac{r^2}{4L_\textup{IR}^2}\biggr)\biggr],
\end{multline}
where ${}_0F_2$ is a generalised hypergeometric function. This Newtonian potential is equivalent to that of GR at short distances $r\ll L_\textup{IR}$, while, for $r\gtrsim L_\textup{IR}$, it is significantly oscillating, signalling a dramatic departure from GR's predictions.

Also in this case, the potentials \eqref{eq:Phi_EUP} and \eqref{eq:Phi_IR} are very different, but the former can be expected to be an effective description of the latter at distances close to $L_\textup{IR}$. The useful equation to be studied is then
\begin{equation}
    \Phi_\textup{EUP}(L_\textup{IR})=\Phi_\textup{NL}(L_\textup{IR}).
\end{equation}
Solving for $\alpha$ leads to
\begin{multline}\label{eq:alpha}
    \alpha=\biggl[\sqrt{\pi}{}_0F_2\biggl(;\frac{1}{2},1;\frac{1}{4}\biggr)-2{}_0F_2\biggl(;\frac{3}{2},\frac{3}{2};\frac{1}{4}\biggr)\\
    -2\sqrt{\pi}\biggr]\frac{L_*^2}{8\sqrt{\pi}G^2m^2}\approx-0.2\,\frac{L_*^2}{G^2m^2}\approx-137\frac{M_\textup{P}^2}{m^2}\frac{L_*^2}{L_\textup{P}^2}.
\end{multline}
As in the UV case, it is worth noticing two important features of this result: $\alpha$ is predicted to be negative and to depend quadratically on the ratio $M_\textup{P}/m$, as well as quadratically on the ratio $L_*/L_\textup{P}$. By substituting this expression for $\alpha$ into that for the event horizon \eqref{eq:R_EH2}, one obtains
\begin{equation}
    R_\textup{EH}=\biggl[{}_0F_2\biggl(;\frac{1}{2},1;\frac{1}{4}\biggr)-\frac{2}{\sqrt{\pi}}{}_0F_2\biggl(;\frac{3}{2},\frac{3}{2};\frac{1}{4}\biggr)\biggr]Gm\approx0.1\,R_\textup{S}.
\end{equation}
The evaporation temperature of the black hole is instead
\begin{equation}
    T_e\approx10\,T_\textup{H}.
\end{equation}
These results point out that both UV and IR effects coming from non-locality could be distinguished from the black hole thermodynamics predicted by GR.
\\

\section{Conclusions}
We compared predictions for static black holes based on modified uncertainty principles and non-local gravity. The key argument we used is that the former can be seen as effective descriptions of the latter at the characteristic length scales where non-local gravitational effects become predominant. The ensuing deviations from GR are due to IDG theories' scrapping of the principle of locality, but the field equations allow to freely attribute these novel consequences to matter too, at least at a phenomenological level. This is why the origin of modified uncertainty principles could ultimately be traced back to the fundamentally non-local nature of gravity. Both in the UV and in the IR regimes we could predict some universal features of GUP- and EUP-inspired black holes respectively: a reduced event horizon compared to the Schwarzschild radius and an increased evaporation temperature compared to the Hawking temperature. Another insightful takeaway is a possible violation of the equivalence principle at the quantum level, which can be expected both from non-local gravity and noncommutative geometry arguments: as the results \eqref{eq:beta} and \eqref{eq:alpha} show, both the parameters $\beta$ and $\alpha$ are predicted to be either directly or inversely proportional to the square of the mass $m$ of the matter source, which would imply that the fundamental commutation relations, i.e.\ respectively the GUP and the EUP, determine a violation of the equivalence principle at appropriate length scales (see also Ref.\ \cite{carr:sub-planckian}). In any case, this result was derived only in the framework of black hole physics, and so its domain of validity remains subject to further investigations in different physical scenarios. In a future work it will also be interesting to study the parameter space of black hole masses under the assumptions that $\beta\sim1$ or $\alpha\sim1$, as well as their respective geodesic structures and observable signatures.

\acknowledgments
S.C.\ and G.M.\ acknowledge the support of Istituto Nazionale di Fisica Nucleare (INFN), Sez.\ di Napoli, Iniziative Specifiche QGSKY and MoonLight-2. This paper is based upon work from COST Action CA21136 -- Addressing observational tensions in cosmology with systematics and fundamental physics (CosmoVerse), supported by COST (European Cooperation in Science and Technology).

\end{document}